\magnification=1180 \vsize=8.9truein \hsize=6.5truein \baselineskip=0.6truecm
\parindent=0truecm \nopagenumbers \font\scap=cmcsc10 \hfuzz=0.8truecm
\parskip=0.2truecm
\font\tenmsb=msbm10
\font\sevenmsb=msbm7
\font\fivemsb=msbm5
\newfam\msbfam
\textfont\msbfam=\tenmsb
\scriptfont\msbfam=\sevenmsb
\scriptscriptfont\msbfam=\fivemsb
\def\Bbb#1{{\fam\msbfam\relax#1}}

\def\qn{q_{n}}
\def\qdo{q_{n-1}}
\def\qup{q_{n+1}}
\def\cn{c_{n}}
\def\cdo{c_{n-1}}

\def\rn{r_{n}}
\def\rup{r_{n+1}}
\def\rdo{r_{n-1}}
\def\rdodo{r_{n-2}}

\def\xup{x_{n+1}}

\def\xn{x_{n}}
\def\yn{y_{n}}
\def\yup{y_{n+1}}

\def\xtup{\tilde{x}_{n+1}}

\def\xtn{\tilde{x}_n}
\def\xt{\tilde{x}}
\def\q{q_{\alpha}}
\def\qu{q_{\alpha+1}}
\def\qd{q_{\alpha-1}}
\def\z{z_{\alpha}}
\def\zu{z_{\alpha+1}}

\def\u{u_{\alpha}}
\def\uu{u_{\alpha+1}}

\def\w{w_{\alpha}}
\def\wu{w_{\alpha+1}}
\def\wd{w_{\alpha-1}}

\newdimen\shirina
\newdimen\glubina
\newdimen\vysota
\newbox\nashbox
\def\underhat#1{\setbox\nashbox=\hbox{$#1$}%
\shirina=\wd\nashbox%
\vysota=\ht\nashbox%
\glubina=\dp\nashbox%
\advance\glubina by .15\vysota%
\hbox to 0pt{\vbox to -\glubina{\hbox to .9\shirina{\hss\^{}}\vss}\hss}#1}

\def\hypotilde#1#2{\vrule depth #1 pt width 0pt{\smash{{\mathop{#2}
\limits_{\displaystyle\tilde{}}}}}}

\def\xdt{\hypotilde 1 x}

\def\ydt{\hypotilde 1 y}

\null \bigskip  \centerline{\bf Discrete systems related to some equations
of the
Painlev\'e-Gambier classification}
\vskip 2cm
\bigskip
\centerline{\scap S. Lafortune$^{\dag}$}
\centerline{\sl LPTM et GMPIB,  Universit\'e Paris VII}
\centerline{\sl Tour 24-14, 5$^e$\'etage}
\centerline{\sl 75251 Paris, France}
\footline{\sl \hskip-1.5cm$^{\dag}$ Permanent address: CRM, Universit\'e de
Montr\'eal, C.P. 6128, Succ.~Centre-ville, Mtl, H3C 3J7, Canada}
\bigskip
\centerline{\scap B. Grammaticos}
\centerline{\sl GMPIB, Universit\'e Paris VII}
\centerline{\sl Tour 24-14, 5$^e$\'etage}
\centerline{\sl 75251 Paris, France}
\bigskip
\centerline{\scap A. Ramani}
\centerline{\sl CPT, Ecole Polytechnique}
\centerline{\sl CNRS, UPR 14}
\centerline{\sl 91128 Palaiseau, France}
\bigskip
\centerline{\scap P. Winternitz}
\centerline{\sl Centre de Recherches Math\'ematiques}
\centerline{\sl Universit\'e de Montr\'eal}
\centerline{\sl C.P. 6128, Succ.~Centre-ville}
\centerline{\sl Montr\'eal, H3C 3J7, Canada}

\vskip 2truecm \noindent Abstract \medskip
\noindent  We derive integrable discrete systems which are contiguity
relations of two equations in the
Painlev\'e-Gambier classification depending on some parameter. These
studies extend earlier work  where the contiguity relations for the six
transcendental Painlev\'e equations were obtained.
In the case of the Gambier equation we give the contiguity relations for
both the continuous and the
discrete system.

\vfill\eject

\footline={\hfill\folio} \pageno=2

\bigskip
\noindent {\scap 1. Introduction}
\medskip
\noindent

The relation of discrete Painlev\'e equations to their, better known,
continuous analogues has by now
been firmly established. As a matter of fact one of the very first methods
proposed to the derivation
of discrete Painlev\'e equations (d-$\Bbb P$) has been the one based on the
auto-B\"acklund/Schlesinger transformations of continuous
Painlev\'e equations [1]. (This link goes even further back in time since
one of the first discrete Painlev\'e equation
 obtained, even before the notion of a d-$\Bbb P$ was explicited, was
the one derived by Jimbo and
Miwa [2] as a contiguity relation of the solutions of
P$_{\rm II}$). The general method for this derivation has been amply
explained in previous publications [3]. One
uses the auto-B\"acklund/Schlesinger transformations in order to derive
relations for the
solutions of a continuous Painlev\'e equations
for contiguous values of the one of its parameters and one gets precisely a
d-${\Bbb P}$.
The discrete equation related to P$_{\rm II}$ has been first obtained, as
explained above,
by Jimbo and Miwa [2] and rediscovered in [1]. The d-${\Bbb P}$'s related
to P$_{\rm III}$ were first
given in [1] and have been recently reexamined in detail in [4]. The
discrete equation related to
P$_{\rm IV}$ was first obtained in [1] and was again studied in [5]. Some
d-${\Bbb P}$'s related to
P$_{\rm V}$ were first given in [1] but the relation of P$_{\rm V}$ to
asymmetric d-P$_{\rm II}$ is
under preparation [6]. Finally the discrete systems obtained from the
auto-B\"acklund of P$_{\rm VI}$
were presented in full detail in [7].
Thus the discrete, difference, Painlev\'e equations can be interpreted
as the contiguity relations of the continuous ones. (This
would tend to cast a doubt on the fundamental
character of the d-$\Bbb P$'s. While this is justified as far as the
difference d-$\Bbb
P$'s are concerned there exists a second type of d-$\Bbb P$'s which have no
relation to
continuous Painlev\'e equations whatsoever. These are the
multiplicative equations, $q$-$\Bbb P$'s [8] which are purely discrete objects
and
thus of
the most fundamental nature). One of the
consequences of this relation between continuous and discrete Painlev\'e
equations is that one can obtain the Lax pairs
of the latter  from the mere existence of the Lax pairs and Schlesinger
transformations of the former. The only
complication comes from the fact that given a d-$\Bbb P$ one has to
identify the procedure for its
derivation based on the Schlesinger transformation of some continuous
Painlev\'e
equation. This can be done in the
frame of the geometric description of Painlev\'e equations which was
introduced in [9] and dubbed
the ``Grand Scheme''. This description based, on the geometry of the affine
Weyl group weight
spaces, has turned out to be a most valuable tool for the classification of
the d-$\Bbb P$'s, which
minimised their uncontrollable proliferation.

In the approach we have described above the emphasis was on the discrete
analogues of the six transcendental equations
of Painlev\'e. As a matter of fact the remaining integrable ODE's of the
Painlev\'e-Gambier
classification have received little attention. Our current knowledge is
essentially limited to the
discrete analogues of the 24 basic equations of Gambier [10]. In particular no
results exist on the
contiguity relations other than for the six equations of Painlev\'e. In
this paper we
intend to focus on just this
question and present a first, exploratory, study of some discrete equations
related to continuous systems
which are not one of the six Painlev\'e equations. In particular we shall
obtain the mappings related to
the equations $35$ and
$27$ of the Painlev\'e-Gambier classification. Equation $27$ is also known
as the Gambier
equation and has been studied in detail in [11], where its discrete equivalent
has been obtained. Since,
as was shown in [12], both the continuous and discrete Gambier equations
possess Schlesinger
transformations, one can obtain the contiguity relations not only of the
continuous but also of the
discrete system. This result is presented in the last part of the present
paper.

\bigskip
\noindent {\scap 2. Discrete systems related to P$_{35}$}
\medskip
\noindent

The first system we are going to examine is the equation \#35 in the
Painlev\'e-Gambier classification [13]:
$$
w''={2\over 3w}w'^2-\left({2\over 3}w-{2\over 3}q-{r\over w}\right)w'+
{2\over 3} w^3-{10\over 3} q w^2+\left(4q'+r+{8\over 3}q^2\right) w+2qr-3r'-
{3r^2\over w}
\eqno(2.1)
$$
with $r=-z/3-2(q'+q^2)/3$ and $q$ one particular solution of
{P}$_{\rm II}$,
$$
v''=2v^3+zv+\alpha,
\eqno(2.2)
$$
where $z$ is the independent variable and $\alpha$ is a constant. The way
to integrate this equation is based on the following Miura transformation.
We start with
the {\sl general} solution of {P}$_{\rm II}$ $v$, and we construct the
quantity
$$
w={v'-q' \over v-q}+v+q
\eqno(2.3)
$$
which solves (2.1). Moreover, starting from the general solution of (2.1)
$w$, the quantity
$$
v={1 \over 3w} (w'+w^2-q w-2 q'-2 q^2-z)
\eqno(2.4)
$$
solves {P}$_{\rm II}$. Equations (2.3) and (2.4) define a Miura
transformation relating
{P}$_{\rm II}$ and P$_{35}$. Another such Miura relation between
P$_{\rm II}$ and some other equation
of the Painlev\'e-Gambier list is already known, that relating P$_{\rm II}$ and
P$_{34}$. It has been studied
in detail in [14] in both the continuous and the discrete case.

Equation P$_{\rm 35}$ has a parameter, $\alpha$, through its relation to
{P}$_{\rm II}$. Let us write
$w_\alpha$  for the
solution of (2.1) corresponding to the parameter $\alpha$ and
$v_\alpha$ and $q_\alpha$ for the solutions of {P}$_{\rm II}$
entering in (2.1), (2.3) and (2.4). Next we consider the auto-B\"acklund
transformations of {P}$_{\rm II}$ [15]:
$$
\eqalignno{
&v_{\alpha+1}=-v_{\alpha}-{\alpha+1/2 \over
v'_{\alpha}+{v_{\alpha}}^2+z/2},&(2.5a)\cr
&v_{\alpha-1}=-v_{\alpha}+{\alpha-1/2 \over
v'_{\alpha}-{v_{\alpha}}^2-z/2},&(2.5b)
}$$
corresponding to solutions where the parameter $\alpha$ varies by integer
quantities. Thus, following the general
theory presented in [1]
one can introduce a discrete system associated to these transformations as a
mapping
which is the contiguity relation for the solutions of {P}$_{\rm II}$.
This equation was in fact first obtained in [2] and is now
known (at least to the present authors) under the name of alternate
d-{P}$_{\rm I}$. It has the form
$$
{\alpha-1/2 \over v_{\alpha-1}+v_\alpha}+{\alpha+1/2\over
v_{\alpha+1}+v_\alpha}=
-2{v_{\alpha}}^2-z. \eqno(2.6)
$$
This is the equation satisfied by all solutions of {P}$_{\rm II}$, and in
particular by $q$, when we
vary the parameter  of
{P}$_{\rm II}$ using (2.5).

Using the Miura (2.3) we can obtain the
solutions of P$_{35}$, $w_{\alpha+1}$ and $w_{\alpha-1}$, corresponding to
the parameter values
$\alpha\pm 1$:
$$
\eqalignno{
&w_{\alpha+1}={v'_{\alpha+1}-q'_{\alpha+1} \over v_{\alpha+1}-q_{\alpha+1}}
+v_{\alpha+1}+q_{\alpha+1},
&(2.7a)
\cr
&w_{\alpha-1}={v'_{\alpha-1}-q'_{\alpha-1} \over v_{\alpha-1}-q_{\alpha-1}}
+v_{\alpha-1}+q_{\alpha-1}.
&(2.7b)
}$$
Using (2.2), (2.3), (2.5) and (2.7), one can write an equation relating
$w_{\alpha+1}$, $w_{\alpha}$ and $w_{\alpha-1}$ alone where neither $v$ nor
its derivatives appear.
Interpreting
$\alpha$ as a discrete independent variable, we obtain thus a second-order
discrete equation for $w$.
This mapping is however  quadratic
in $w_{\alpha+1}$ and $w_{\alpha-1}$ and thus (as was also argued in [16])
cannot be integrable. Our argument is based on the fact that the evolution
of such a mapping leads in
general to an exponential number of images, and preimages, of the initial
point. The
non-integrability of this non-singlevalued system is not in contradiction to
the fact that we can obtain one  solution of this mapping, namely the one
furnished by the evolution of
the system before elimination. This solution is the only one that we know
how to describe, while a
system with exponentially increasing number of branches eludes a full
description.

In order to proceed, instead of working with a second-order mapping, we
choose to derive a system of two
coupled equations.
We first make use of the discrete
symmetry of the solutions of {P}$_{\rm II}$: if $v_\alpha$ is a solution
of
{P}$_{\rm II}$ with parameter $\alpha$, then $-v_\alpha$ is a solution of
{P}$_{\rm II}$ with parameter $-\alpha$. We then introduce the quantity
$$
u_{\alpha}={v'_{\alpha}-q'_{\alpha} \over v_{\alpha}-q_{\alpha}}
-v_{\alpha}-q_{\alpha}
\eqno(2.8)
$$
which is obtained from the expression (2.3) for $w_\alpha$ after implementing
the reflections $v_\alpha \rightarrow -v_\alpha$ and $q_\alpha \rightarrow
-q_\alpha$. The quantity $u_\alpha$ is thus the solution of a {\sl slightly
modified}
P$_{35}$ equation (with $-q_\alpha$ instead of $q_\alpha$) that we shall denote
$\tilde{\rm P}_{\rm 35}$. Note that, from (2.3) and (2.8) we have:
$$
w_\alpha - u_\alpha=2(v_\alpha + q_\alpha).
\eqno(2.9)
$$

We start by writing an equation between $w_{\alpha-1}$, $w_\alpha$ and
$u_{\alpha}$
only,
using (2.2), (2.3), (2.5), (2.7) and (2.9).
Similarly we also obtain an equation relating
$u_{\alpha+1}$, $u_\alpha$ and $w_{\alpha}$. We thus find the
second-order discrete system:
$$
\eqalignno{
&\uu=\w+ {2(\qu+\q)^2\w^2 \over
(\qu+\q)(\u\w-\w^2+2(\q-\qu)\w)+2\alpha+1},&(2.10a)\cr
&\w=\u+{1-2\alpha\over \u(\q+\qd)}+{2\wd(\q-\qd)-4\q\u \over \wd-\u}, &(2.10b)
}
$$
where $q_\alpha$ is a particular solution of (2.6).

This system introduces the contiguity relations for equation P$_{35}$ (2.1)
and
$\tilde{\rm P}_{35}$.  It can
be integrated with the help of alternate d-{P}$_{\rm I}$, equation (2.6). We
first write
$v'_\alpha$ in terms of $v_{\alpha+1}$ and $v_\alpha$ and $q'_{\alpha}$ in
terms of $q_{\alpha+1}$ and $q_{\alpha}$ using the Schlesinger transformation
(2.5). Using these expressions in the definition
(2.3), we obtain the following:
$$
w_\alpha = {\alpha +1/2 \over v_\alpha-q_\alpha}
\left[ {1\over q_\alpha +q_{\alpha+1}}-{1\over v_\alpha +v_{\alpha+1}}\right].
\eqno(2.11)
$$
The quantity $u_\alpha$ is obtained in terms of $v_\alpha$ and $w_\alpha$
using (2.9).
Thus, once $v_\alpha$ is known, we can construct $w_\alpha$ and $u_\alpha$ in
a straigthforward way.

At this point we can remark that while (2.10) is integrable by
construction, the condition
for integrability, namely that $q$ is a specific function, may appear
somewhat contrived. In what
follows we shall show that the fact that $q$
satisfies alternate d-P$_{\rm I}$ can
be obtained precisely through the singularity confinement criterion. We
start by writing (2.10) and (2.6) in a more convenient way in order to
implement the singularity
analysis.:
$$
\eqalignno{
&\uu=\w+{2(\qu+\q)^2\w^2\over (\qu+\q)(\w\u-\w^2+2(\q-\qu)\w)+
2\zu},&(2.12a)\cr
&\w=\u-{2\z\over \u(\q+\qd)}+{2\wd(\q-\qd)-4\q\u\over \wd-\u},&(2.12b)
}$$
where $q$ and $z$ are two functions to be determined.

There are three possible sources of singularities for (2.12): $\u=0$,
$\w=0$ and the case when the
denominator in the rhs of equation (2.12a) happens to be 0 for some value of
$\alpha$. We first examine the
singularity
$\u=0$. Let the initial  condition $\wd$ be free and introduce the small
parameter $\epsilon$:
$\u=\epsilon$. At the lowest order in $\epsilon$ we find, $\w=1/\epsilon+
\dots$, $\uu=1/\epsilon+\dots$, $\wu=\cal{O}(\epsilon)$.  Then
${u_{\alpha+2}}$ must take the
indefinite form $0/0$ in order to contain the information on $\wd$ when
$\epsilon\rightarrow 0$.
Implementing this condition, we find that $q_\alpha$ must indeed satisfy
$$
{\z\over \qd+\q }+{\zu\over \qu+\q }=-2{\q}^2-\gamma,
\eqno(2.13)
$$
for some constant $\gamma$ and
$\z$ must be linear in $\alpha$. (The two other singularities are
automatically confined
for all functions $q_\alpha$ and $z_\alpha$.) Thus the singularity
confinement criterion
gives for
$q_\alpha$ and $z_\alpha$ precisely the constraints expected from the
construction of
(2.12) from P$_{\rm 35}$.

\bigskip
\noindent {\scap 3. A mapping related to the continuous Gambier
equation}
\medskip
\noindent

The second discrete system we are going to derive is the contiguity
relation of the solutions of the
Gambier equation. The continuous Gambier system can be written as a cascade
of two Riccati  equations:
$$
y'=-y^2+c,
\eqno(3.1a)
$$
$$
x'=ax^2+nxy+\sigma,
\eqno(3.1b)
$$
where $a$, $c$ and $\sigma$ are free functions of the independent variable $z$
and $n$ is an integer. By a change of variable, the function
$\sigma$ can
be
scaled to $0$ or $1$ but, for the purpose of constructing Schlesinger
transformations,
it is more convenient to keep it as a free function. Moreover, if we
require that
equation
(3.1) possess the Painlev\'e property, the functions will be subject to
one more
constraint.
This constraint depends on the value of the integer parameter $n$.

Let us write $x_n$, $a_n$ and $\sigma_n$, the quantities appearing
in (3.1) corresponding to the coupling parameter $n$ in (3.1b). Then, we can
introduce
the following Schlesinger transformation [12,17]. It relates
the solutions of
(3.1) to that of an equation of the same form but with a coupling
parameter
$\tilde{n}=n+2$. This transformation is given by:
$$
\eqalignno{
&x_{n+2}=-{\sigma_{n+2}\over n+1}\left[y+{1\over n+2}{a_n'\over a_n}+{a_n
\xn\over n+1}
\right]^{-1}, & (3.2a) \cr
&{\sigma'_{n+2} \over \sigma_{n+2}}=-{n\over n+2}{a_n'\over a_n}, & (3.2b)
\cr
&a_{n+2}={n+1\over \sigma_{n+2}}\left(\cn+{a_n\sigma_n \over n+1}
+{1\over n+2}{a_n''\over a_n}-{n+3\over (n+2)^2}{a_n'^2\over a_n^2}\right).
&(3.2c)
}$$
The inverse of this transformation is given by
$$
\eqalignno{
&x_{n-2}={1-n\over a_{n-2}}\left[{\sigma_n\over x_n(n-1)}+y+{1\over 2-n}
{\sigma_n'\over \sigma_n}\right],&(3.3a)\cr
&{a'_{n-2} \over a_{n-2}}=-{n\over n-2}{\sigma_n'\over \sigma_n},& (3.3b)\cr
&\sigma_{n-2}={1-n\over a_{n-2}}\left(c_n+{a_n\sigma_n \over 1-n}
+{1\over 2-n}{\sigma_n''\over \sigma_n}+{n-3\over (n-2)^2}{\sigma_n'^2\over
\sigma_n^2}\right). &(3.3c)
}$$
We can obtain a second-order mapping for $x$
eliminating $y$ between equations (3.2a) and (3.3a).
$$
x_{n-2}  {a_{n-2}\over n-1}  -
 \xn   {a_n\over n+1}-   {\sigma_n'\over \sigma_n}{1\over n-2}
- {a_n'\over a_n}{1\over n+2} +{1\over \xn} {\sigma_n\over n-1}
-{1\over x_{n+2}}{\sigma_{n+2}\over n+1}=0\eqno(3.4)
$$
There is another way to present equation (3.4) if we write it as a system,
involving
explicitly $y$ and using the fact that it does not depend on $n$. The two
equations (3.2b) and (3.2c)
are equivalent to the upshifts (by 2 units of $n$) of (3.3b) and (3.3c).
Equation (3.2b)
is the compatibility condition between (3.2a) and the upshift of (3.3a).

We obtain thus
$$
\eqalignno{
&y_{n+2}=y_n&(3.5a)\cr
&x_{n-2}={1-n\over a_{n-2}}\left[{\sigma_n\over x_n(n-1)}+y_n+{1\over 2-n}
{\sigma_n'\over \sigma_n}\right]. &(3.5b)
}$$
The evolution of $\sigma$ and $a$ is obtained through equations
(3.2b)-(3.2c). System (3.5) is just a
particular case of the Gambier mapping introduced in [18]. It is
remarkable that the contiguity
relation of the solutions of the Gambier equations with respect to the one
integer parameter, namely $n$,
is again a (discrete) Gambier equation. However the latter does not have
the full freedom of the
Gambier mapping (in perfect parallel to the relation between continous
Painlev\'e equations and
d-P's.)

\bigskip
\noindent {\scap 4. A mapping related to the discrete Gambier
equation}
\medskip
\noindent

In this section we shall concentrate on the generic discrete version of the
Gambier equation. As was
shown in [12] it can be written as a  system
of two discrete Riccati's in cascade:
$$
\eqalignno{
&\yup={\yn+\cn\over \yn+1},
&(4.1a)\cr
&\xup={\xn(\yn-\rn)+\qn(\yn-s_n)\over \xn\yn}.
&(4.1b)
}$$
(Other choices for the
Gambier mapping do exist but equation (4.1) is the most convenient for the
derivation of the Schlesinger
transformation.)
The mapping was analysed in [17] using the singularity confinement method. The
only singularity that plays a role is that induced by a special value of
$y$, which,
given the form of (4.1), is $y=0$. The way for this singularity to be
confined is to
require that the value of $x$ go through an indeterminate value $0/0$ after
a certain
number of steps, say $N$. This means in particular that $x$ must be zero,
while $s$
appearing in (4.1b) must be equal to the quantity
$\psi_N$ (introduced in [12]) which is the $N$th iterate of $y=0$ under (4.1a),
$N$ times
downshifted i.e.
$\psi_{0,n}=0$,
$\psi_{1,n}=c_{n-1}$,
$\psi_{2,n}={c_{n-2}+c_{n-1} \over
c_{n-2} +1}$, etc.
As was shown in [12], at the continuous limit, the number $N$ of steps,
necessary for
confinement, goes over to the parameter $n$ appearing in the continuous Gambier
equation (3.1).

The discrete version of the Schlesinger transformation (3.2) is a
transformation which
relates the solution of a mapping confining in $N$ steps to that of a
mapping confining
in $N+2$ steps. The complete analysis for the derivation of the discrete
Schlesinger
can be found in [12]. Here we shall just state the result:
$$
\xtup=\phi_n{\xn\rn+\qn\psi_{N,n}\over \yn}{\yn-\psi_{N+1,n}\over
\psi_{N,n-1}(\xn-1)+\rdo},
\eqno(4.2)
$$
where $\xt$ is the solution requiring $N+2$ steps for confinement and $x$
stands for
the solution necessitating just $N$ steps. The quantity $\phi$ is given by
$$
\phi_n={(\psi_{N,n-1}+1)(\psi_{N,n-1}\psi_{N,n-2}\qdo
+\psi_{N,n-2}r_{n-1}-r_{n-1}\,r_{n-2})
\over
(\psi_{N,n-2}-\rdodo)(\psi_{N,n}
\qn+\rn\rdo+\rn)+
\rn\qdo\,\psi_{N,n-2}(1+\psi_{N,n-1})}.
\eqno(4.3)
$$
With this proper choice of $\phi$ we find that $\xt$ satisfies an equation:
$$
\xtup={\xtn(\yn-\tilde{r}_n)+\tilde{q}_n(\yn-\psi_{N+2,n})\over \xtn \yn},
\eqno(4.4)
$$
where $\tilde{r}$ and $\tilde{q}$ are functions completely determined in
terms of
$q$, $r$ and $c$. In a similar way we can derive the Schlesinger transformation
invoving a solution which takes $N-2$ steps to confine. We find:
$$
{\xdt}_n=\omega_n{\xn(\yn-\cdo)+g_n(\yn-\psi_{N,n})\over
\xn(\yn-\cdo)+h_n(\yn-\psi_{N,n})},
\eqno(4.5)
$$
where
$$
\eqalignno{
&g_n={\qup(\psi_{N-1,n+1}-\cn)\over \psi_{N-1,n+1}-
\rup},&(4.6a)\cr
&h_n={({\cdo}-{\rn})({\cn}-1)\over
{\cdo}+{\cn}-\psi_{N,n+1}({\cdo}+1)},
&(4.6b)\cr
&\omega_n={u_n(1-{\cn})+
h_n(1+{\cn}-2\psi_{N,n+1})\over u_n(1-{\cn})+
g_n(1+{\cn}-2\psi_{N,n+1})
},&(4.6c)\cr
&u_n=1-\rn+{\qn(\cdo-1)\over h_{n-1}}.
&(4.6d)
}$$
We can now obtain the contiguity relation of the solutions of the Gambier
mapping by
eliminating $y$ between (4.2) and (4.5).

On the other hand, just as in the case of the continuous Gambier system
there exists a
simpler way to write this mapping. The value of $y$ appearing in (4.2) or
(4.5) is
independent of $N$. Thus a far simpler way to present the mapping is for
instance as
$$
\eqalignno{
&\ydt_n =\yn, & (4.7a)\cr
&\xdt_n=\omega_n {\xn(\yn-\cdo)+g_n(\yn-\psi_{N,n})\over
\xn(\yn-\cdo)+h_n(\yn-\psi_{N,n})}.&(4.7b) }
$$
Again we remark that (4.7) is a mapping of Gambier type although of a very
particular
one.

\bigskip
\noindent {\scap 5. Conclusion}
\medskip

In this paper we have studied the contiguity relations of the solutions of
two equations in
the Painlev\'e-Gambier classification which contain some parameter. This
kind of
investigation was previously limited to the six transcendental equations of
Painlev\'e.
The main result of those studies was to show that the discrete Painlev\'e
equations are
just the contiguity relations of the continuous Painlev\'e equations. Here
we have
extended this approach to equations which, while not being one of the six
Painlev\'e
equations are integrable and, in fact, belong to the same classification.
In the case of
the Gambier equation we have used the Schlesinger transformation which is
associated to the integer parameter
$n$ appearing in the equation. Our result in the case of the continuous Gambier
equation is that the contiguity relation of the solutions satisfies a
mapping which
assumes the form of a discrete analogue of the Gambier system, in a nice
analogy to the
situation for the continuous and discrete Painlev\'e equations. In parallel
to our approach for d-${\Bbb P}$'s the discrete Gambier system has also been
analysed. We have
shown that using the Schlesinger transformation that also exist in this
case one can derive the contiguity relation of the solutions which turns
out to be in the form of
(a special case of) a Gambier mapping. We expect our approach to be
applicable to
other equations
of the Painlev\'e-Gambier classification which contain parameters and for which
Miura/Schlesinger transformations exist.
Now that the question of classification of discrete Painlev\'e equations is
in the process of being
settled [19,20] it is interesting to analyse the remaining equations in the
classification and work out
their implications in the discrete case.
\bigskip
\noindent {\scap Acknowledgments}
\medskip
\noindent S. L. acknowledges two scholarships: one from FCAR for his Ph.D.
and one from ``Le Programme
de Soutien de Cotutelle de Th\`ese de doctorat du Gouvernement du Qu\'ebec''
for his stay in Paris. The research of P.W. was partly supported by
research grant from
NSERC of Canada and FCAR du Qu\'ebec. S. L. would like to thank the CNRS
for their hospitality.
B.G and A.R. acknowledge illuminating e-correspondence with P.A. Clarkson.
\bigskip
\noindent {\scap References}.
\medskip
\item{[1]} A.S. Fokas, B. Grammaticos and A. Ramani, J. Math. Anal. Appl.
180  (1993), 342.
\item{[2]} M. Jimbo and T. Miwa, Physica D  2 (1981), 407.
\item{[3]} B. Grammaticos, F. Nijhoff and A. Ramani, {\sl Discrete Painlev\'e
equations}, in: {\sl The Painlev\'e Property: One Century Later} (ed. R.
Conte), CRM Series in Mathematical Physics, pp 413-516, Springer-Verlag,
New-York (1999).
\item{[4]} P.A. Clarkson, E.L. Mansfield and H.N. Webster, {\sl On discrete
Painlev\'e equations as
B\"acklund transformations}, to appear in the proceedings of CRM-AARMS
Workshop on B\"acklund and
Darboux transformations, (ed. D. Levi and P. Winternitz), Halifax, June 1999.
\item{[5]} B. Grammaticos and A. Ramani,  J. Phys. A 31 (1998) 5787.
\item{[6]} B. Grammaticos, A. Ramani and T. Tokihiro,  in preparation.
\item{[7]} F. Nijhoff, B. Grammaticos, A. Ramani and Y. Ohta, {\sl On
discrete Painlev\'e equations
 associated with the lattice KdV systems and the Painlev\'e VI equation},
to appear in Stud. Appl. Math.
\item{[8]} A. Ramani, B. Grammaticos and J. Hietarinta, Phys. Rev. Lett.
67 (1991), 1829.
\item{[9]} A. Ramani, Y. Ohta, J. Satsuma, B. Grammaticos, Comm. Math.
Phys.  192 (1998), 67.
\item{[10]} B. Grammaticos and A. Ramani, Meth. Appl. Anal.  4 (1997), 196.
\item{[11]} B. Grammaticos and A. Ramani, Physica A  223 (1996), 125.
\item{[12]} A. Ramani, B. Grammaticos and S. Lafortune, Lett. Math. Phys.,
46 (1998), 131.
\item{[13]} E. L. Ince, {\sl Ordinary differential equations}, Dover, New
York, 1956.
\item{[14]} A. Ramani and B. Grammaticos, J. Phys. A  25 (1992), L633.
\item{[15]} A.S. Fokas and M.J. Ablowitz, J. Math. Phys.  23 (1982), 2033.
\item{[16]} M.D. Kruskal, K.M. Tamizhmani, B. Grammaticos and A. Ramani, {\sl
Asymmetric Discrete Painlev\'e Equations}, Preprint 1999.
\item{[17]} B. Gambier, Acta Math. 33 (1910) 1.
\item{[18]} B. Grammaticos, A. Ramani and S. Lafortune, Physica A  253
(1998), 260.
\item{[19]} H. Sakai, {\sl Rational surfaces associated with affine root
systems and geometry of the
Painlev\'e equations}, PhD thesis, Kyoto University (1999).
\item{[20]} B. Grammaticos and A. Ramani, Reg. Chaot. Dyn. 5 (2000) 53.

\end